\documentclass[]{tMPH2e}

\usepackage{amsmath}

    \usepackage{bm}
    \usepackage{graphicx}

\newcommand\beq{\begin{equation}}
\newcommand\eeq{\end{equation}}
\newcommand\beqa{\begin{eqnarray}}
\newcommand\eeqa{\end{eqnarray}}

\def\etal{\mbox{et al.}}

\begin{document}

\markboth{Santos \etal}{Contact values for disparate-size hard-sphere mixtures}


\title{Contact values for disparate-size hard-sphere mixtures}

\author{Andr\'es Santos\thanks{$^\ast$Email: andres@unex.es}$^\ast$, Santos B. Yuste\thanks{$^\dagger$ Email:
santos@unex.es}$^\dagger$ and Mariano L\'{o}pez de Haro\thanks{$^\ddagger$ On sabbatical leave from Centro de Investigaci\'on en
Energ\'{\i}a, Universidad Nacional Aut\'onoma de M\'exico
(U.N.A.M.), Temixco, Morelos 62580, M{e}xico. Email:
malopez@servidor.unam.mx}$^\ddagger$
 \\\vspace{6pt}
 Departamento de F\'{\i}sica, Universidad de Extremadura, E-06071
Badajoz, Spain\\
\vspace{6pt}
Morad Alawneh\thanks{$^\S$Email: morad.alawneh@gmail.com}$^\S$ and
Douglas Henderson\thanks{$^\P$Email: doug@chem.byu.edu}$^\P$
\\\vspace{6pt}
Department of Chemistry and Biochemistry, Brigham Young
University, Provo, UT 84602-5700, U.S.A.}

\received{\today}

\maketitle

\begin{abstract}
A universality ansatz for the contact values of a multicomponent
mixture of additive hard spheres is used to propose new formulae for
the case of disparate-size binary mixtures. A comparison with
simulation data and with a recent proposal by Alawneh and Henderson
for binary mixtures shows reasonably good agreement with the
predictions for the contact values of the large-large radial
distribution functions. A discussion on the usefulness and
limitations of the new proposals is also presented.
\end{abstract}


\section{Introduction}
A well known result in statistical mechanics is the close
relationship between the structural and thermodynamic properties of
fluids. In the particular case of a multicomponent mixture of
(additive) hard spheres (HS), the compressibility factor $Z\equiv
p/\rho k_B T$ (where $p$, $\rho$ and $T$ are the pressure, density
and temperature of the system and $k_B$ is the Boltzmann constant)
is given by

\begin{equation}
Z=1+4\eta\sum_{i,j}^N\frac{\sigma_{ij}^3}{\langle
\sigma^3\rangle}x_i x_j g_{ij}(\sigma_{ij}),
\label{1EOS}
\end{equation}
where $N$ is the number of components in the mixture, $x_{i}$ is the
mole fraction of species $i$, $\sigma_{ij}=\frac{1}{2 }(\sigma
_{i}+\sigma _{j})$, $\sigma _{i}$ being the diameter of a sphere of
species $i$, $\eta =(\pi/6) \rho \langle \sigma^3\rangle$ is the
packing fraction (with $\langle \sigma^n\rangle\equiv\sum_{i=1}^N
x_{i}\sigma _{i}^{n}$) and $g_{ij}(\sigma_{ij})$ are the contact
values  of the radial distribution functions (RDFs) $g_{ij}(r)$.
Hence, in this system, knowledge of the contact values of the RDFs
suffices to completely determine the equation of state (EOS). Exact
results for $g_{ij}(\sigma_{ij})$ are unfortunately only known to
first order in the packing fraction, where one simply has

\begin{equation}
g_{ij}(\sigma_{ij})=1+\eta \left(1+\frac{3}{2} z_{ij}\right) +
\emph{O}(\eta^2),
\label{gij_eta}
\end{equation}
where
\begin{equation}
z_{ij}\equiv \frac{\sigma _{i}\sigma_{j}}{\sigma _{ij}}\frac{\langle
\sigma^2\rangle}{\langle \sigma^3\rangle}.
\label{zij}
\end{equation}

Therefore, in general one must rely on either approximations or
values obtained from computer simulations. Amongst the most
important approximate analytical expressions for the contact values
are the ones that follow from the solution of the Percus--Yevick
(PY) equation of additive HS mixtures by Lebowitz \cite{L64}, namely
\beq
g_{ij}^{\text{PY}}(\sigma_{ij})=\frac{1}{1-\eta
}+\frac{3}{2}\frac{\eta }{(1-\eta )^{2}}z_{ij},
\label{15PY}
\eeq
and the results obtained from the Scaled Particle Theory (SPT)
\cite{RFL59,MR75,R88,HC04},
\beq
g_{ij}^{\text{SPT}}(\sigma_{ij})=\frac{1}{1-\eta
}+\frac{3}{2}\frac{\eta }{(1-\eta
)^{2}}z_{ij}+\frac{3}{4}\frac{\eta^{2}}{(1-\eta)^{3}}z_{ij}^{2}.
\label{15SPT}
\eeq

However, neither the PY nor the SPT lead to particularly accurate
values and so Boubl\'{\i}k \cite{B70} and, independently, Grundke
and Henderson \cite{GH72} and Lee and Levesque \cite{LL73} proposed
an interpolation between the PY and the SPT contact values, that we
will refer to as the BGHLL values:
\beq
g_{ij}^{\text{BGHLL}}(\sigma_{ij})=\frac{1}{1-\eta
}+\frac{3}{2}\frac{\eta }{(1-\eta
)^{2}}z_{ij}+\frac{1}{2}\frac{\eta^{2}}{(1-\eta)^{3}}z_{ij}^{2}.
\label{15BGHLL}
\eeq
These lead, through Eq.\ (\ref{1EOS}), to the popular and widely
used Boubl\'{\i}k--Mansoori--Carnahan--Starling--Leland (BMCSL) EOS
\cite{B70,MCSL71} for HS mixtures. For binary mixtures the BGHLL
values are rather accurate except for asymmetric mixtures in which
large hard spheres are present in extreme dilution in a small
hard-sphere solvent. Since such mixtures represent good models for
colloidal suspensions and there is growing interest in these kinds of
systems, there have been many efforts to correct the deficiencies of
the BGHLL values
\cite{HMLC96,YCH96,ML97,HC98,HBCW98,MHC99,CCHW00,BS00,VS02,HTWC05,VT06,TR07,AH08}.
In this paper the recent proposal of Ref.\ \cite{AH08} is of
particular concern, so we also quote it explicitly here. It relies
heavily on the proposal of Viduna and Smith (VS) \cite{VS02} which
reads
\beq
g_{ij}^{\text{VS}}(\sigma_{ij})=\frac{1}{1-\eta}+\eta\frac{3-\eta+\eta^2/2}{2(1-\eta)^2}z_{ij}+\eta^2\frac{2-\eta-\eta^2/2}{3(1-\eta)^3}
\left(1+\frac{\langle\sigma\rangle\langle\sigma^3\rangle}{2\langle\sigma^2\rangle^2}\right)z_{ij}^2.
\label{20}
\eeq

Alawneh and Henderson (AH) \cite{AH08} restrict themselves to binary
mixtures ($N=2$, $\sigma_2 \ge \sigma_1$) and propose to keep VS's
expression for $g_{11}(\sigma_{1})$ but amend $g_{12}(\sigma_{12})$
and $g_{22}(\sigma_{2})$ as
\beq
g_{12}^{\text{AH}}(\sigma_{12})=g_{12}^{\text{VS}}(\sigma_{12})-\frac{\eta^3}{16(1-\eta)^3}(1-R^{-3})z_{12}^3
+\frac{5\eta^4}{32(1-\eta)^3}(1-R^{-4})z_{12}^4-\frac{\eta^5}{16(1-\eta)^3}(1-R^{-5})z_{12}^5,
\label{21}
\eeq
\beq
g_{22}^{\text{AH}}(\sigma_{2})=g_{22}^{\text{VS}}(\sigma_{2})+\frac{3\eta}{2}(z_{22}-1)\frac{1-R^{-1}}{R}\exp\left[\frac{3\eta}{2}(z_{22}+1)\right].
\label{22}
\eeq
Here $R\equiv \sigma_{2}/\sigma_{1}$ \cite{note}. With these, they
get good agreement for the large-small contact values and reasonably
good agreement for the large-large contact values as compared to
Molecular Dynamics results for mixtures with size ratios going from
$R=1$ to $R=10$.

On a different vein, in recent papers \cite{SYH02,SYH05,SYH06,HYS08}
three of us have advocated a ``universality'' approach as a guide to
propose simple and accurate approximate expressions for the contact
values $g_{ij}(\sigma_{ij})$ of  additive HS mixtures. In this
approach, if the packing fraction $\eta$ is fixed, the expressions for the
contact values of the RDF for all pairs $ij$ of like and unlike
species depend on the diameters of both species and on the
composition of the mixture \textit{only} through the single
dimensionless parameter $z_{ij}$ [cf. Eq.\ \eqref{zij}],
irrespective of the number of components in the mixture. Note that
this feature is shared by the PY, SPT and BGHLL results but not by
the VS or the AH approximations. The aim of this paper is two-fold.
On the one hand, we will assess whether the new simulation data of
Ref.\ \cite{AH08} comply with the universality ansatz. On the other
hand, we will introduce a refinement of some of the simple proposals
of Refs.\ \cite{SYH02,SYH05,SYH06,HYS08} that retain the
universality feature but are able to improve the performance in the
case of highly asymmetric mixtures.

The paper is organized as follows. In Sect.\ \ref{sec2} and in order
to make the paper self-contained, we provide a rather brief account
of the simple proposals based on the universality ansatz and a
simple extension to deal with highly asymmetric systems. This is
followed in Sect.\ \ref{sec3} by a comparison between the AH contact
values, those of the new proposals and the simulation data. The
paper is closed in Sect.\ \ref{sec4} with further discussion and
some concluding remarks.

\section{Proposals for contact values including the ``universality'' assumption}
\label{sec2}

We start by recalling the main aspects of the ``universality''
approach that has been used recently \cite{SYH02,SYH05,SYH06,HYS08}
as a guide to propose simple and accurate approximate expressions
for the contact values $g_{ij}(\sigma_{ij})$ of additive HS
mixtures. According to this approach, at a given total packing
fraction $\eta$, $g_{ij}(\sigma_{ij})$ depends on the size and
concentration distributions only through the single parameter
$z_{ij}$:
\begin{equation}
g_{ij}(\sigma _{ij})=G(\eta,z_{ij}),
\label{5}
\end{equation}
where the function $G(\eta,z)$ is  a common function for all the
pairs $(i,j)$, regardless of the {composition and} number of
components of the mixture. Note that the exact contact values to
first order in density [cf. Eq.\ \eqref{gij_eta}] are consistent
with the ansatz \eqref{5} with $G(\eta,z) \to 1 +
\eta\left(1+\frac{3}{2}z\right)$.

In order to propose specific forms for $G(\eta,z)$  use is made of
three basic consistency conditions. First, in the limit in which one
of the species, say $i$, is made of point particles ({i.e.}, $\sigma
_{i}\rightarrow 0$), $g_{ii}(\sigma _{i})$ must take the ideal gas
value, except that one has to take into account that the available
volume fraction is $1-\eta$. Thus,
\begin{equation}
\lim_{\sigma_{i}\rightarrow 0}g_{ii}(\sigma_{i})=\frac{1}{1-\eta}.
\label{2}
\end{equation}
Next, if all the species have the same size, $\{\sigma
_{k}\}\rightarrow \sigma $,  the system becomes equivalent to a
single component system, so that
\begin{equation}
\lim_{\{ \sigma _{k}\}\rightarrow \sigma
}g_{ij}(\sigma_{ij})=g_{p}(\eta), \label{3}
\end{equation}
where $g_{p}(\eta)$ is the contact value of the RDF of the single
component fluid at the same packing fraction $\eta$ as that of the
mixture.

The third condition is the equality between the pressure in the bulk
and near a hard wall \cite{SYH06,HYS08}:
\beq
1+4\eta\sum_{i,j}\frac{\sigma_{ij}^3}{\langle \sigma^3\rangle}x_i
x_j g_{ij}(\sigma_{ij})=\sum_j x_j g_{wj},
\label{wi}
\eeq
where $g_{wj}$ is the wall-particle correlation function at contact.
Since a hard wall can be seen as a sphere of infinite diameter
\cite{HAB76}, the contact value $g_{wj}$ can be obtained from
$g_{ij}(\sigma_{ij})$ as
\beq
g_{wj}=\lim_{\stackrel{\sigma_{i}\to\infty}{
x_{i}\sigma_{i}^3\rightarrow 0}} g_{ij}(\sigma_{ij}).
\label{3p}
\eeq
In the special case of a pure fluid plus a hard wall, Eq.\
\eqref{wi} becomes
\beq
\lim_{\stackrel{\sigma_{2}\to\infty}{ x_{2}\sigma_{2}^3\rightarrow
0}} g_{12}(\sigma_{12})=1+4\eta g_p(\eta)\equiv Z_p(\eta),
\label{3pb}
\eeq
where $Z_p(\eta)$ is the compressibility factor of the pure fluid.

Assuming the validity of the ansatz \eqref{5}, conditions \eqref{2},
\eqref{3}, \eqref{wi} and \eqref{3pb} become
\beq
G(\eta,0)=\frac{1}{1-\eta},
\label{1a}
\eeq
\beq
G(\eta,1)=g_p(\eta),
\label{2ab}
\eeq
\beq
1+4\eta\sum_{i,j}\frac{\sigma_{ij}^3}{\langle \sigma^3\rangle}x_i
x_j G(\eta,z_{ij})=\sum_j x_j G(\eta,z_{wj}),\quad z_{wj}\equiv
2\sigma_{j}\frac{\langle\sigma^2\rangle}{\langle\sigma^3\rangle},
\label{zwi}
\eeq
\beq
G(\eta,2)=Z_p(\eta),
\label{3ab}
\eeq
respectively. Conditions \eqref{1a}, \eqref{2ab} and \eqref{3ab}
are fulfilled by the quadratic proposal \cite{SYH02,HYS08}
\beq
G_{e2}(\eta,z)=\frac{1}{1-\eta}+\left[g_p(\eta)-\frac{1}{1-\eta}\right] z
+\left[\frac{1-\eta/2}{1-\eta}-(1-2\eta)g_p(\eta)\right]z(z-1).
\label{6}
\eeq
However, $G_{e2}$ does not satisfy Eq.\ \eqref{zwi} for an arbitrary
number of components $N \ge 3$. This is accomplished by a cubic form
\cite{SYH05,SYH06,HYS08}:
\beq
G_{e3}(\eta,z)=G_{e2}(\eta,z)+(1-\eta)\left[g_p^{\text{SPT}}(\eta)-g_p(\eta)\right]z(z-1)(z-2),
\label{7}
\eeq
where
\beq
g_p^{\text{SPT}}(\eta)=\frac{1-\eta/2+\eta^2/4}{(1-\eta)^3}
\label{8}
\eeq
is the SPT contact value of the pure fluid. The proposals $G_{e2}$
and $G_{e3}$ are very flexible in the sense that they can
accommodate any reasonable choice for $g_p(\eta)$. In particular if
$g_p(\eta)=g_p^{\text{SPT}}(\eta)$ then
$G_{e2}(\eta,z)=G_{e3}(\eta,z)$ and one recovers the SPT contact
values given in Eq.\ (\ref{15SPT}).

Taking either the  $g_p(\eta)$ values that follow from the
Carnahan--Starling or the Carnahan--Starling--Kolafa (CSK) EOS as
input, both approximations $G_{e2}$ and $G_{e3}$ yield in general
very good predictions for $g_{ij}(\sigma_{ij})$ in binary and
polydisperse mixtures under a wide range of conditions
\cite{SYH02,SYH05,SYH06,HYS08}, which typically implies $z_{ij}\leq
2$. However, in the case of very disparate mixtures, the parameter
$z_{ij}$ can take larger values. For instance, in a binary mixture
with $R>1$ one has
\beq
z_{22}(x_2,R)=R\frac{1-x_2+x_2 R^2}{1-x_2+x_2 R^3}.
\label{17}
\eeq

\begin{figure}
\includegraphics[width=.7 \columnwidth]{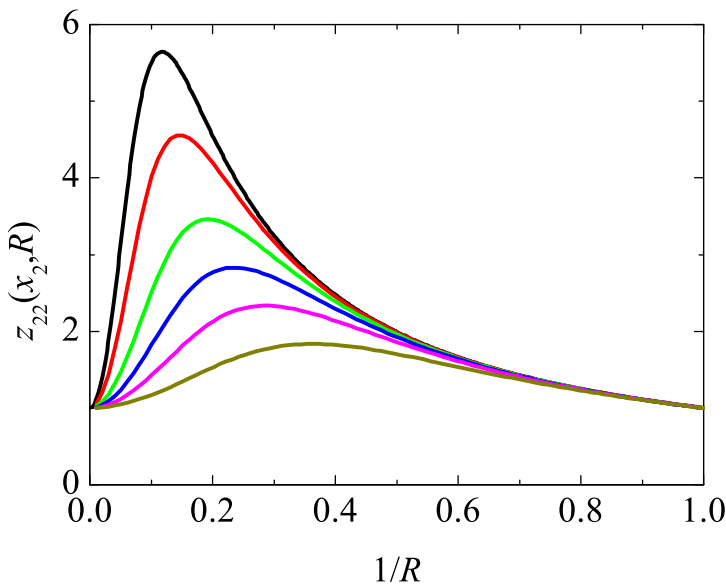}
\caption{(Colour online) Plot of $z_{22}(x_2)$  as a function of
$1/R$ for, from top to bottom,  $x_2=0.001$, $0.002$, $0.005$,
$0.01$, $0.02$ and $0.05$.
\label{zmax}}
\end{figure}
\begin{figure}
\includegraphics[width=.7 \columnwidth]{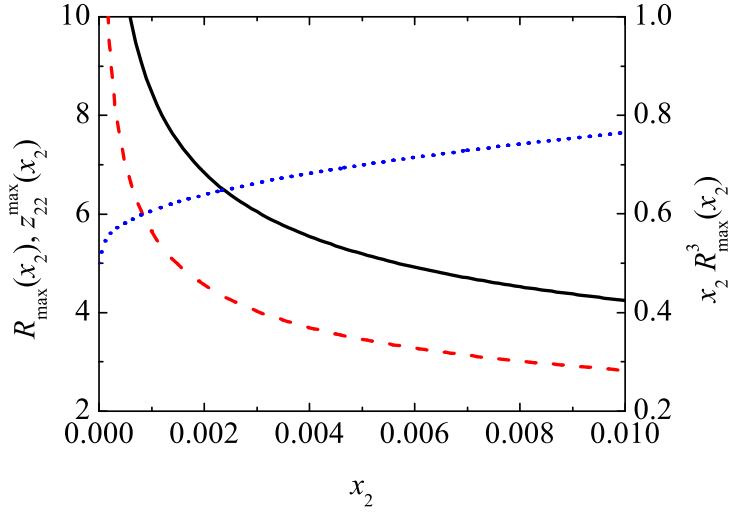}
\caption{(Colour online) Plot of $R_\text{max}(x_2)$ (solid line) and
the corresponding $z_{22}^\text{max}(x_2)$ (dashed line). Also shown
is the behaviour of the quantity $x_2 R_\text{max}^3$ (dotted line, right vertical
scale) as a function of $x_2$.
\label{Rmax}}
\end{figure}

The behaviour of $z_{22}$ as a function of $1/R$ is shown in Fig.
\ref{zmax} for different values of $x_2$. It follows that, at fixed
$x_2$, there exists a value $R=R_\text{max}(x_2)$ at which $z_{22}$
reaches a maximum $z_{22}^\text{max}(x_2)$. The peak
$z_{22}^\text{max}(x_2)$ becomes narrower and higher as $x_2$
decreases, its location moving to the left (larger values of $R$).
An explicit expression for $R_\text{max}(x_2)$ can be derived by
differentiation of $z_{22}$ with respect to $R$, which yields a
cubic equation for $R$ whose real solution is
\beq
R_\text{max}(x_2)=\frac{1}{2}\left[1+\left(2x_2^{-1}+2x_2^{-1}\sqrt{1-x_2}-1\right)^{1/3}+\left(2x_2^{-1}+2x_2^{-1}\sqrt{1-x_2}-1\right)^{-1/3}\right].
\label{18a}
\eeq

A plot of $R_\text{max}(x_2)$ and the associated values of
$z_{22}^\text{max}(x_2)$ is shown in Fig.\ \ref{Rmax}. In the limit
$x_2\to 0$ one gets $R_{\text{max}}\to (2x_2)^{-1/3}$, so that
$\lim_{x_2\to 0} x_2 R_{\text{max}}^3=1/2$. For $x_2>0$ one has $x_2
R_{\text{max}}^3>1/2$. This means that, when $z_{22}$ reaches its
maximum value, the partial volume fraction occupied by the big
spheres is always larger than half the one occupied by the small
ones. Figure \ref{Rmax} shows that $z_{22}^\text{max}(x_2)$ can
reach rather high values for sufficiently small $x_2$. In
particular, $z_{22}^\text{max}(x_2)>2$ if $x_2\leq 0.036$,
irrespective of the value of $R$. Since $G(\eta,z)$ is a
monotonically increasing function of $z$, the larger $z_{22}$, the
higher $g_{22}(\sigma_{2})$. Thus, Eq.\ \eqref{18a} gives the value
of $R$ at which $g_{22}(\sigma_{2})$ has a maximum, according to
the ansatz \eqref{5}. Note that $R^\text{max}(x_2)$ is independent
of $\eta$.

The polynomial forms \eqref{6} and \eqref{7} are constructed by
imposing exact properties at $z=0$, $z=1$ and $z=2$. On the other
hand, in disparate mixtures where one can have values of $z$ higher
than $2$, Eqs.\ \eqref{6} and \eqref{7} are not accurate enough. In
fact, again in the binary case, there exists evidence indicating
that $g_{22}(\sigma_{2})$ grows exponentially for large $R$, as
explicitly incorporated in the AH proposal \cite{AH08}, Eq.\
(\ref{22}). This translates, in the context of the ansatz \eqref{5},
into an exponential dependence of $G(\eta,z)$ on $z$ for large $z$.
Thus, here we propose to amend Eqs.\ \eqref{6} and \eqref{7} as
\beq
G_{e2+}(\eta,z)=\begin{cases} G_{e2}(\eta,z),&0\leq z\leq 2,\\
Z_p(\eta)\exp\left[A_{e2}(\eta)(z-2)\right],&2\leq z,
\end{cases}
\label{18}
\eeq
\beq
G_{e3+}(\eta,z)=\begin{cases} G_{e3}(\eta,z),&0\leq z\leq 2,\\
Z_p(\eta)\exp\left[A_{e3}(\eta)(z-2)\right],&2\leq z,
\end{cases}
\label{18b}
\eeq
where $A_{e2}(\eta)$ and $A_{e3}(\eta)$ are determined by requiring
continuity of the first derivative at the matching point $z=2$.
Their expressions are
\beq
A_{e2}(\eta)= \frac{1}{Z_p(\eta)}\left[\frac{2-3
\eta/2}{1-\eta}-2(1-3 \eta) g_p(\eta)\right],
\label{19a}
\eeq
\beq
A_{e3}(\eta)= \frac{1}{Z_p(\eta)}\left[\frac{4-9
\eta/2+2\eta^2}{(1-\eta)^2}-4(1- 2 \eta) g_p(\eta)\right].
\label{19b}
\eeq

Note that the amended proposals \eqref{18} and \eqref{18b} also
share with Eqs.\ \eqref{6} and \eqref{7} the flexibility to
accommodate any reasonable choice for $g_p$. In what follows we will
use the one corresponding to the CSK EOS, namely
\beq
g_p^{\text{CSK}}(\eta)=\dfrac{1-\eta /2+\eta^2(1-2\eta)/12}{(1-\eta
)^{3}}.
\label{gpCSK}
\eeq

\section{Comparison with simulation data}
\label{sec3}
\begin{figure}
\includegraphics[width=.7 \columnwidth]{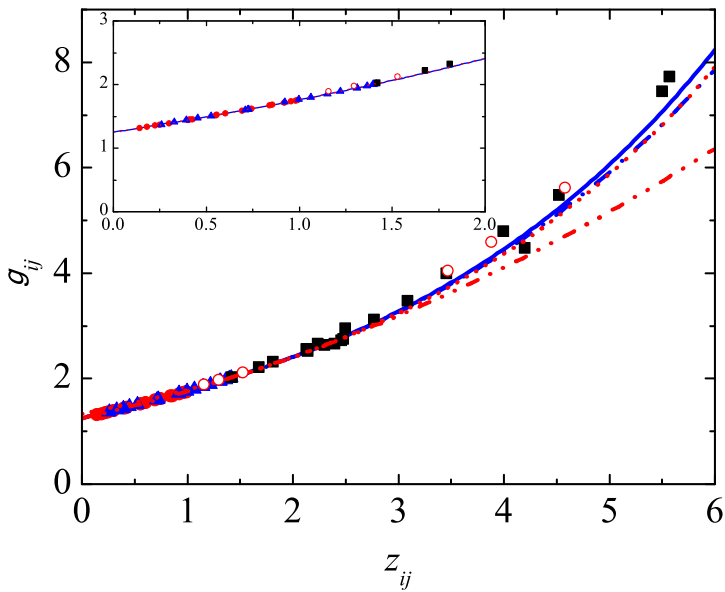}
\caption{(Colour online) Plot of $g_{ij}(\sigma_{ij})$ as a function
of $z_{ij}$ for $\eta=0.2$. Dash-dot-dot line: $G_{e2}(\eta,z)$;
dash-dot line: $G_{e3}(\eta,z)$; dotted line: $G_{e2+}(\eta,z)$;
solid line:  $G_{e3+}(\eta,z)$; solid circles: simulation data of
$g_{11}(\sigma_{1})$ from Ref.\ \cite{AH08};  solid triangles:
simulation data of $g_{12}(\sigma_{12})$ from Ref.\
\cite{AH08};  solid squares: simulation data of
$g_{22}(\sigma_{2})$ from Ref.\ \cite{AH08};   open circles:
simulation data of $g_{w1}$ and $g_{w2}$ from Ref.\
\cite{MYSH07}. The inset shows the region $0\le z_{ij} \le 2$.
Note that $G_{e2+}(\eta,z)$ and $G_{e3}(\eta,z)$ are practically
indistinguishable at this density.\label{eta02}}
\end{figure}
\begin{figure}
\includegraphics[width=.7 \columnwidth]{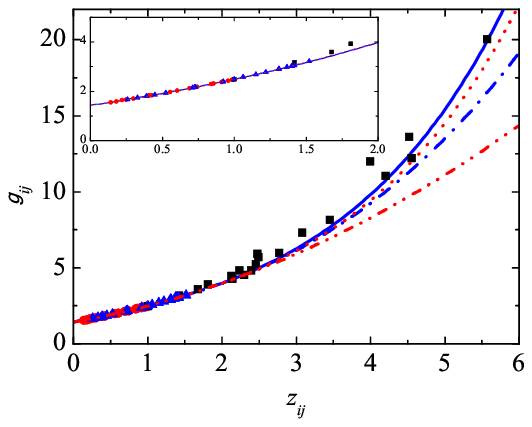}
\caption{(Colour online) Same as in Fig.\ \protect\ref{eta02}, but
for $\eta=0.3$. At this density there are no simulation data for
$g_{wj}$.
\label{eta03}}
\end{figure}
\begin{figure}
\includegraphics[width=.7 \columnwidth]{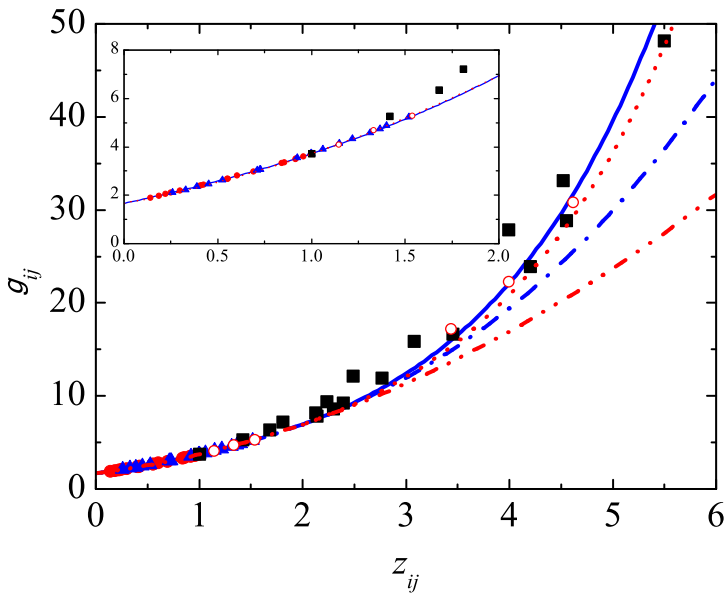}
\caption{(Colour online) Same as in Fig.\ \protect\ref{eta02}, but
for $\eta=0.4$.\label{eta04}}
\end{figure}
\begin{figure}
\includegraphics[width=.7 \columnwidth]{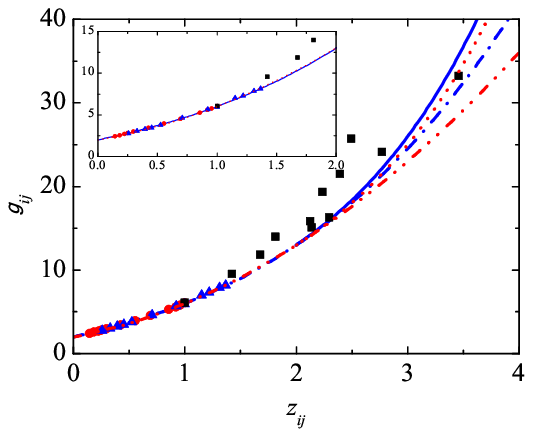}
\caption{(Colour online) Same as in Fig.\ \protect\ref{eta02}, but
for $\eta=0.5$. At this density there are no simulation data for
$g_{wj}$.\label{eta05}}
\end{figure}

Once the new universal proposals have been introduced, the question
now arises as to whether the available simulation data are
consistent with the universality ansatz. Were this to be the case,
for a given packing fraction $\eta$ plots of $g_{ij}$ vs $z_{ij}$ of
different mixtures should lie on a common curve and this will be
tested next.

The approximations $G_{e2}(\eta,z)$, $G_{e3}(\eta,z)$,
$G_{e2+}(\eta,z)$ and $G_{e3+}(\eta,z)$,  given in Eqs.\ \eqref{6}
\eqref{7}, \eqref{18} and \eqref{18b}, respectively, complemented
with the CSK expression for $g_p(\eta)$ given in Eq.\ \eqref{gpCSK},
are compared with simulation data of $g_{ij}(\sigma_{ij})$
\cite{AH08,MYSH07} in Figs.\ \ref{eta02}--\ref{eta05}. We can
observe that $G_{e2}$ and $G_{e3}$ perform very well for $0<z<2$,
except for the higher values of $g_{22}(\sigma_{2})$ in that region
(see the insets of Figs.\ \ref{eta02}--\ref{eta05}). Since, by
construction, both $G_{e2}$ and $G_{e3}$ give the correct value at
$z=2$ [cf. Eq.\ \eqref{3ab}], the deviations of the simulation
values of $g_{22}(\sigma_{2})$ from the theoretical curves in the
region $0<z<2$ can be interpreted as a manifestation of the
approximate character of the universality ansatz \eqref{5}. This
seems to be confirmed by the scatter of the simulation data of
$g_{22}(\sigma_{2})$ for $z>2$. On the other hand, the overall
behaviour of $g_{22}(\sigma_{2})$ in that region seems to be
reasonably well captured by both $G_{e2+}$ and $G_{e3+}$.

\begin{figure}
\includegraphics[width=.7 \columnwidth]{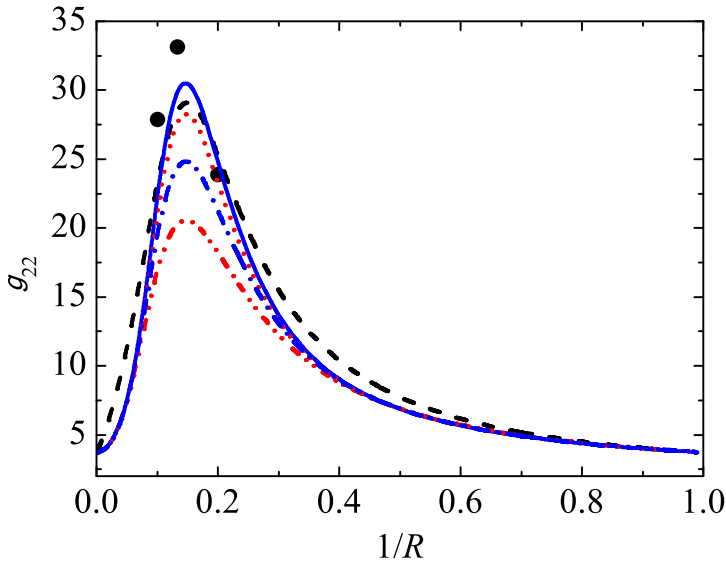}
\caption{(Colour online) Plot of $g_{22}(\sigma_{2})$ as a function
of $1/R$ for $\eta=0.4$ and $x_2=0.002$ (as given by different
approximations) and simulation results from Ref.\ \cite{AH08}. Dash-dot-dot line: $g_{22}^{e2}(\sigma_2)$; dash-dot line: $g_{22}^{e3}(\sigma_2)$; dotted line: $g_{22}^{e2+}(\sigma_2)$; solid line: $g_{22}^{e3+}(\sigma_2)$;  dashed line:
$g_{22}^{\text{AH}}(\sigma_{2})$; solid circles: simulation
data of $g_{22}(\sigma_{2})$ from Ref.\ \cite{AH08}.
\label{Fig.1b}}
\end{figure}
\begin{figure}
\includegraphics[width=.7 \columnwidth]{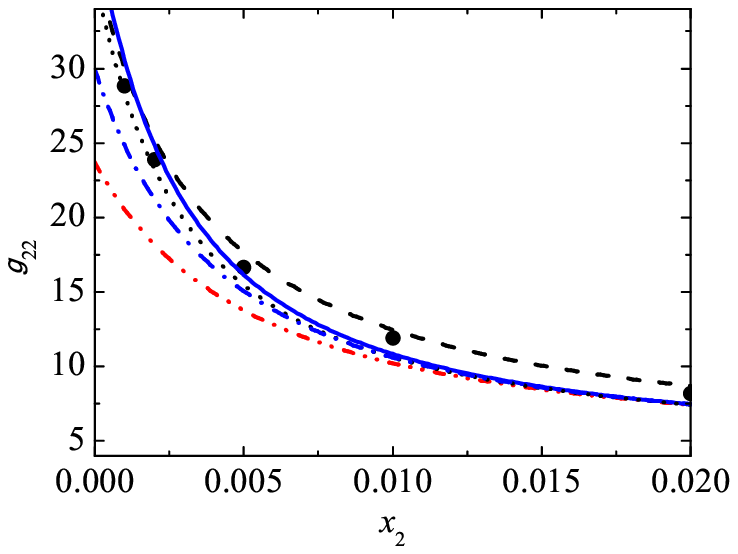}
\caption{(Colour online) Plot of $g_{22}(\sigma_{2})$ as a function
of $x_2$ for $\eta=0.4$ and $R=5$ (as given by different
approximations) and simulation results from Ref.\ \cite{AH08}.  Dash-dot-dot line: $g_{22}^{e2}(\sigma_2)$; dash-dot line: $g_{22}^{e3}(\sigma_2)$; dotted line: $g_{22}^{e2+}(\sigma_2)$; solid line: $g_{22}^{e3+}(\sigma_2)$;  dashed line:
$g_{22}^{\text{AH}}(\sigma_{2})$; solid circles: simulation
data of $g_{22}(\sigma_{2})$ from Ref.\ \cite{AH08}.
\label{Fig.S9c}}
\end{figure}
\begin{figure}
\includegraphics[width=.7 \columnwidth]{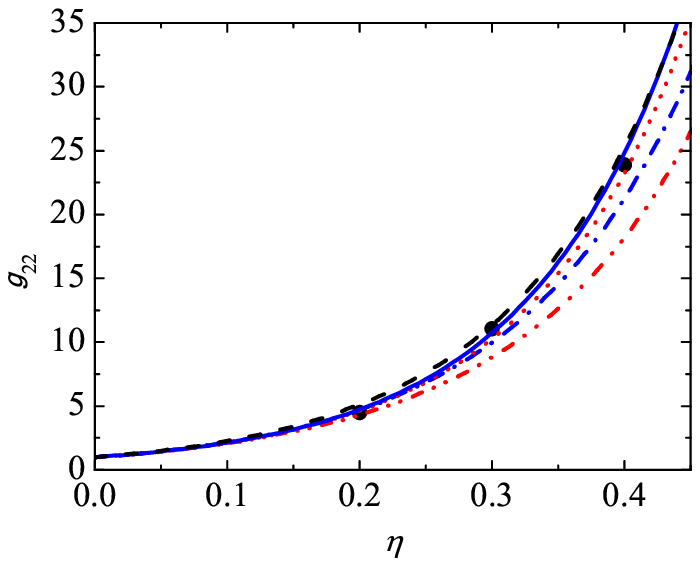}
\caption{(Colour online)  Plot of $g_{22}(\sigma_{2})$ as a function
of $\eta$ for $x_2=0.002$ and $R=5$ (as given by different
approximations) and simulation results from Ref.\ \cite{AH08}. Dash-dot-dot line: $g_{22}^{e2}(\sigma_2)$; dash-dot line: $g_{22}^{e3}(\sigma_2)$; dotted line: $g_{22}^{e2+}(\sigma_2)$; solid line: $g_{22}^{e3+}(\sigma_2)$;  dashed line:
$g_{22}^{\text{AH}}(\sigma_{2})$; solid circles: simulation
data of $g_{22}(\sigma_{2})$ from Ref.\ \cite{AH08}.
\label{Fig.MA1}}
\end{figure}
\begin{figure}
\includegraphics[width=.7 \columnwidth]{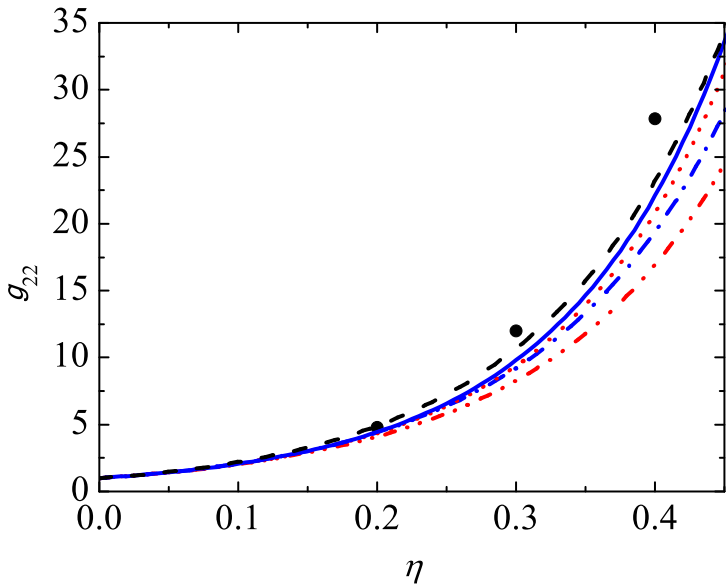}
\caption{(Colour online) Same as in Fig.\ \protect\ref{Fig.MA1}, but for $x_2=0.002$ and $R=10$.
\label{Fig.MA3}}
\end{figure}

Once the question of universality has been examined, in order to
complement the perspective of the present results, in Figs.\
\ref{Fig.1b}--\ref{Fig.MA3} we illustrate some other issues. In
particular, we want to assess the  performance of the theoretical
approximations obtained with the universality assumption, denoted by
$g_{22}^*(\sigma_{2})$ (where the asterisk will be either $e2$,
$e3$, $e2+$ or $e3+$ depending on whether one uses $G_{e2}(\eta,z)$,
$G_{e3}(\eta,z)$, $G_{e2+}(\eta,z)$ or $G_{e3+}(\eta,z)$, respectively, to compute
the contact value) and that of the AH approximation
$g_{22}^{\text{AH}}(\sigma_{2})$ [cf. Eq.\ \eqref{22}] as compared
to simulation data. Hence, in Fig.\ \ref{Fig.1b} we have plotted
$g_{22}(\sigma_{2})$ as a function of $1/R$ taking $x_2=0.002$ and
$\eta=0.4$, while in Fig.\ \ref{Fig.S9c} the large-large contact
value has been plotted as a function of $x_2$ for $\eta=0.4$ and
$R=5$. These plots suggest that the new approximations may indeed
improve the already reasonable agreement between the AH formula and
simulation results, but all theories seem to underestimate the
height of the peak in the $g_{22}(\sigma_{2})$ vs. $1/R$ plot. The
region of interest, namely relatively large $R$ and small $x_2$, is
further examined in Figs.\ \ref{Fig.MA1} and \ref{Fig.MA3} which
illustrate two additional points. On the one hand, one can clearly
see that the curves are almost indistinguishable for low densities
(up to $\eta \simeq 0.2$ in these cases) but at higher densities the
new proposals $G_{e2+}(\eta,z)$ and $G_{e3+}(\eta,z)$ indeed improve
on the performance of the earlier ones $G_{e2}(\eta,z)$ and
$G_{e3}(\eta,z)$. On the other hand, they also show that, in the
high density region, the AH formula performs a little bit better
than either $G_{e2+}(\eta,z)$ or $G_{e3+}(\eta,z)$ as $R$ becomes
larger.

\section{Concluding remarks}
\label{sec4}

The above results deserve some further consideration. In this paper,
following the universality approach that some of us have followed
for some years, we have introduced two new proposals [cf. Eqs.\
\eqref{18} and \eqref{18b}] for the contact values of HS mixtures
which improve on the earlier proposals in the case where one has
disparate in size mixtures. These proposals fulfill certain exact
conditions and are rather flexible, requiring as their only input
the contact value of the one-component system.

The comparison with recent computer simulation data
\cite{AH08,MYSH07} indicates that the universality  assumption may
have some limitations but, since there are many technical
difficulties in simulating mixtures with large $R$ and very small
$x_2$, one cannot reach definite conclusions on this issue on the
basis of these limited data. On the other hand, the new proposals
following the universality assumptions for the large-large contact
values seem nevertheless to yield reasonably good agreement both
with the simulation results and with the recent proposal by Alawneh
and Henderson \cite{AH08}.

It is clear that the AH formulae do not obey the ansatz \eqref{5}.
First, the non-universal character of Eq.\ \eqref{20} is due to the
appearance of the combination
$\langle\sigma\rangle\langle\sigma^3\rangle/{\langle\sigma^2\rangle^2}$.
In addition, the values given by Eqs.\ \eqref{21} and \eqref{22}
depend on $1-R^{-n}$. As mentioned above, our analysis of the
universality issue in the light of the simulation results suggests
that the ansatz \eqref{5}  may break down at some point and hence
whether the AH proposal is universal or not would not be
particularly relevant. However, a more serious problem with the AH
formulae is that Eq.\ \eqref{22} fails to reduce to the exact result
to first order in $\eta$ given in Eq.\ \eqref{gij_eta}. In contrast,
all the universal approximations quoted in this paper as well as the
VS approximation yield such result in the appropriate limit. Whether
or not the universality feature is confirmed or discarded depends on
the availability of further simulation results. Our hope is that
this paper may encourage the performance of such simulations.

\section*{Acknowledgements}
The work of A. S., S. B. Y. and M. L. H. has been supported by the
Ministerio de Educaci\'on y Ciencia (Spain) through Grant No.
FIS2007-60977 (partially financed by FEDER funds) and by the Junta
de Extremadura through Grant No. GRU09038.



\begin{thebibliography}{99}

\bibitem{L64}
{{J.~L.} {Lebowitz}},
  {Phys. Rev. A} \textbf{{133}},
  {895} ({1964}).

\bibitem{RFL59}
H. Reiss, H. L. Frisch and J. L. Lebowitz, J. Chem. Phys.
\textbf{31}, 369 (1959); E. Helfand, H. L. Frisch and J. L.
Lebowitz, J. Chem. Phys. \textbf{34}, 1037 (1961); J. L. Lebowitz,
E. Helfand and E. Praestgaard, J. Chem. Phys. \textbf{43}, 774
(1965).

\bibitem{MR75}
 M. J. Mandell  and H. Reiss, J. Stat. Phys.  \textbf{13}, 113 (1975).

\bibitem{R88}
{{Y.}~{Rosenfeld}},
  {J. Chem. Phys.} \textbf{{89}},
  {4272} ({1988}).


\bibitem{HC04}
M. Heying and D. S. Corti, J. Phys. Chem. B \textbf{108}, 19756
(2004).

\bibitem{B70}
{{T.}~{Boubl\'{\i}k}},
  {J. Chem. Phys.} \textbf{{53}},
  {471} ({1970}).

\bibitem{GH72}
{{E.~W.} {Grundke}} {and}
  {{D.}~{Henderson}},
  {Mol. Phys.} \textbf{{24}},
  {269} ({1972}).

\bibitem{LL73}
{{L.~L.} {Lee}} {and}
  {{D.}~{Levesque}},
  {Mol. Phys.} \textbf{{26}},
  {1351} ({1973}).

\bibitem{MCSL71}
{{G.~A.} {Mansoori}},
  {{N.~F.} {Carnahan}},
  {{K.~E.} {Starling}}
  {and}
  {{J.}~{T.~W.~Leland}},
  {J. Chem. Phys.} \textbf{{54}},
  {1523} ({1971}).

\bibitem{HMLC96}
{{D.}~{Henderson}},
  {{A.}~{Malijevsk\'{y}}},
  {{S.}~{Lab\'{\i}k}}
  {and} {{K.~Y.} {Chan}},
  {Mol. Phys.} \textbf{{87}},
  {273} ({1996}).

\bibitem{YCH96}
{{D.~H.~L.} {Yau}},
  {{K.-Y.} {Chan}} {and}
  {{D.}~{Henderson}},
  {Mol. Phys.} \textbf{{88}},
  {1237} ({1996}); \textbf{91}, 1137 (1997).

\bibitem{ML97}
D. V. Matyushov and B. M. Ladanyi, J. Chem. Phys. \textbf{107}, 5815
(1997).

\bibitem{HC98}
{{D.}~{Henderson}} {and}
  {{K.~Y.} {Chan}},
  {J. Chem. Phys.} \textbf{{108}},
  {9946} ({1998});   Mol. Phys. \textbf{94}, 253
  (1998); \textbf{98}, 1005 (2000).

\bibitem{HBCW98}
{{D.}~{Henderson}},
  {{D.}~{Boda}},
  {{K.~Y.} {Chan}} {and}
  {{D.~T.} {Wasan}},
  {Mol. Phys.} \textbf{{95}},
  {131} ({1998}).

  \bibitem{MHC99}
D. Matyushov, D. Henderson and K.-Y. Chan, Mol. Phys. \textbf{96},
1813 (1999).

\bibitem{CCHW00}
{{D.}~{Cao}},
  {{K.-Y.} {Chan}},
  {{D.}~{Henderson}}
  {and} {{W.}~{Wang}},
  {Mol. Phys.} \textbf{{98}},
  {619} ({2000}).


\bibitem{BS00}
{{C.}~{Barrio}} {and}
  {{J.~R.} {Solana}},
  {J. Chem. Phys.} \textbf{{113}},
  {10180} ({2000}).

\bibitem{VS02}
D. Viduna and W. R. Smith, Mol. Phys. \textbf{100}, 2903 (2002); J.
Chem. Phys. \textbf{117}, 1214 (2002).

\bibitem{HTWC05}
D. Henderson, A. Trokhymchuk, L. V. Woodcock and K.-Y. Chan, Mol.
Phys. \textbf{103}, 667 (2005).

\bibitem{VT06}
A. Vrabecz and G. T\'oth, Mol. Phys. \textbf{104}  1843 (2006).

\bibitem{TR07}
A. Takhtoukh and C. Regnaut, Fluid Phase Equil. \textbf{262}, 149 (2007).

\bibitem{AH08}
M. Alawneh and D. Henderson, Mol. Phys. \textbf{106}, 607 (2008).

\bibitem{note}
Note the typographical error in Eq.\ (7) of Ref.
\protect\cite{AH08} where a factor of $R$ is missing in the
prefactor of the exponential correction to the VS formula.

\bibitem{SYH02}
A. Santos, S. B. Yuste and M. L\'opez de Haro, J. Chem. Phys.
\textbf{117}, 5785 (2002).

\bibitem{SYH05}
A. Santos, S. B. Yuste and M. L\'opez de Haro, J. Chem. Phys.
\textbf{123}, 234512 (2005).

\bibitem{SYH06}
 M. L\'opez de Haro, S. B. Yuste and A.
Santos,  Mol. Phys. \textbf{104}, 3461 (2006).

\bibitem{HYS08}
M. L\'opez de Haro, S. B. Yuste and A. Santos, \emph{Alternative
Approaches to the Equilibrium Properties of Hard-Sphere Liquids}, in
\emph{Theory and Simulation of Hard-Sphere Fluids and Related
Systems}, Lectures Notes in Physics, vol. 753, A. Mulero, ed.
(Springer, Berlin, 2008), pp. 183--245.


\bibitem{HAB76}
D. Henderson, F. F. Abraham and J. A. Barker, Mol. Phys.
\textbf{31}, 1291 (1976).

\bibitem{MYSH07}
Al. Malijevsk\'y, S. B. Yuste, A. Santos and M. L\'opez de Haro,
 Phys. Rev. E \textbf{75}, 061201 (2007).




\end{thebibliography}
\end{document}